\documentclass[lettersize,journal]{IEEEtran}
\usepackage{amsmath,amsfonts}
\usepackage{algorithmic}
\usepackage{algorithm}
\usepackage{array}
\usepackage{textcomp}
\usepackage{stfloats}
\usepackage{url}
\usepackage{xcolor}
\usepackage{verbatim}
\usepackage{graphicx}
\usepackage{cite}
\usepackage{subcaption}
\usepackage{caption}
\usepackage{amsmath}
\usepackage{multirow}
\usepackage{booktabs}
\usepackage{adjustbox}
\usepackage{makecell}

\usepackage{graphicx}
\usepackage{subcaption}
\usepackage{ragged2e}

\usepackage{amsthm}
\theoremstyle{remark}

\begin{document}


\title{Dynamic Resilience Assessment of Power Systems With Data Center Load Events Using Physics-Informed Neural Networks}

\author{Chen Chao,~\IEEEmembership{Graduate Student Member,~IEEE};
Zixiao Ma,~\IEEEmembership{Member,~IEEE};
Ziang Zhang,~\IEEEmembership{Senior Member,~IEEE}
\thanks{The authors are with the Department of Electrical and Computer Engineering, Binghamton University, State University of New York, Binghamton, NY 13902 USA (e-mail: cchao3@binghamton.edu; zma10@binghamton.edu; ziang.zhang@binghamton.edu). Corresponding author: Zixiao Ma}
}

\markboth{Journal of \LaTeX\ Class Files,~Vol.~14, No.~8, August~2015}%
{Shell \MakeLowercase{\textit{et al.}}: Bare Demo of IEEEtran.cls for IEEE Journals}

\maketitle

\begin{abstract}
Large data center loads introduce new resilience challenges to power systems because their disconnection and staged reconnection can induce fast voltage and frequency dynamics that are not captured by static service-status or energy-based metrics. This paper proposes a utility-side, physics-informed resilience assessment framework that evaluates these events using only grid-side dynamic models and observable post-disturbance trajectories, without requiring detailed internal data center models. An unsupervised differential algebraic equation-physics informed neural network (DAE-PINN) based on an implicit backward Euler residual is developed to jointly predict dynamic and algebraic states, enabling repeated post-disturbance trajectory evaluation while enforcing network algebraic consistency. Normalized multi-phase resilience metrics are then used to quantify
disturbance, degraded-state, and restoration-period impacts and to screen
data center reconnection timing and load-ramping strategies under security
constraints. Case studies on a modified IEEE 33-bus feeder show that the proposed DAE-PINN accurately tracks numerical DAE solutions and substantially reduces computation time in repeated restoration screening. The proposed metrics distinguish the effects of disturbance size, data center location, and reconnection strategy, revealing the trade-off between restoration speed and transient resilience loss.

\end{abstract}


\IEEEpeerreviewmaketitle

\section{Introduction}

\IEEEPARstart{T}he rapid growth of data centers is reshaping electric power systems\cite{shehabi2024united}. Driven by cloud computing and artificial intelligence workloads, data centers already account for a significant share of electricity demand in the United States, and this trend is expected to continue\cite{aljbour2024powering}. Unlike traditional industrial loads, modern data centers are often large, disturbance-sensitive, and capable of causing abrupt changes in system operating conditions after disconnection. Such events can trigger significant voltage and frequency excursions, as illustrated by the large-load-loss incident in North Virginia\cite{NERC2025LargeLoadLoss}. Therefore, resilience assessment for large data center events should quantify not only whether service is eventually restored, but also how voltage, frequency, and other system states deviate during load disconnection, degraded operation, and staged reconnection.

Power system resilience has been widely studied in recent years, yet its definition and quantification remain application dependent\cite{raoufi2020power}. Many existing metrics are built from aggregate performance curves\cite{vugrin2017resilience}, outage durations\cite{dobson2023long}, and area-based loss measures\cite{dobson2023models}. These metrics are useful for broad event characterization, but they are less suitable for disturbances whose main effect is a fast change in system dynamics after the disconnection of a large load. In such cases, resilience depends on transient state evolution, including deviation magnitude, settling behavior, and time to return to secure operating bounds \cite{panteli2017metrics, stankovic2022methods}. Trajectory-based assessment is therefore a natural basis for resilience quantification in this problem.

Implementing such trajectory-based assessments in practice, however, faces a significant informational challenge: utilities typically do not have access to detailed internal data center models. 
Operational and performance details of data centers are often unavailable because of privacy, communication, and cybersecurity concerns\cite{quint2025practical, NERC2025EmergingLargeLoads}. 
This limitation is especially important because existing data center power and load modeling studies largely rely on information that is internal to the facility.
Recent work has developed short-term load and power forecasting models\cite{mughees2025short},  server-level power models \cite{yu2025enhanced}, and broader reviews of AI data center energy demand and grid impacts \cite{sheng2026power, chen2025electricity}. 
Related studies also formulate data center load and flexibility models for demand response and power-system support \cite{zhou2025data}.
However, these approaches still generally assume access to internal workload characteristics, server operating information, or data center-side coordination variables that are not visible to utilities. 
As a result, utility-side resilience assessment after data center disconnection cannot generally rely on full internal data center models and should instead be based on grid-side dynamic behavior.
In practice, such behavior is characterized by the time-domain trajectories of relevant system variables following a disturbance. 
Therefore, the problem is not only to define trajectory-based resilience metrics, but also to generate post-disturbance trajectories efficiently and in a way that remains consistent with grid-side physics under limited information sharing. This creates a utility-side screening problem: many candidate reconnection and restoration strategies must be evaluated using physics-consistent grid trajectories, while the detailed workload, server-level, and backup-system models inside the data center remain unavailable.

Traditionally, generating these grid-side trajectories requires repeated simulation of nonlinear differential-algebraic equations (DAEs), which can be computationally expensive across many scenarios. Physics-informed neural networks (PINNs) offer a promising alternative by embedding governing equations directly into the learning process\cite{raissi2019physics}. In power-system applications, prior studies have shown that PINNs can approximate system trajectories and estimate parameters in swing-equation models\cite{misyris2020physics, stiasny2021transient}. DAE-PINN formulations in \cite{moya2023dae} further showed that nonlinear differential-algebraic trajectories can be learned while respecting algebraic constraints and stiff behavior, although validation was limited to a trivial 3-bus system. Recent PINN-based frameworks support faster post-disturbance trajectory prediction for operational studies\cite{stiasny2024pinnsim}, while \cite{shen2025physics} developed physics-following neural networks for dynamic security assessment but still relied on simulation or numerical procedures to ensure algebraic consistency. Existing PINN methods for power-system dynamics therefore still face limitations in optimization robustness, algebraic consistency, and computational efficiency when many post-disturbance scenarios must be evaluated.

To address this gap, this paper develops a utility-side resilience
assessment framework for large data center disconnection and reconnection
events. The framework evaluates each candidate disturbance or reconnection
strategy through a physics-consistent grid trajectory and maps the trajectory
to phase-wise resilience and security indicators. To enable repeated
screening, an unsupervised DAE-PINN based on an implicit backward Euler
residual is developed to jointly predict dynamic and algebraic states. The
predicted trajectories are then evaluated by multi-phase metrics and
security constraints. The main contributions are summarized as follows:
\begin{itemize}
\item We formulate a utility-side, trajectory-based resilience assessment problem for large data center disconnection and reconnection events. The formulation does not require internal data center workload or server-level models and instead evaluates grid-facing impacts through post-disturbance voltage, frequency, and angle trajectories.
\item We develop an unsupervised DAE-PINN trajectory predictor based on an implicit backward Euler residual. The model jointly predicts dynamic and algebraic states and explicitly penalizes both differential and algebraic residuals, thereby improving algebraic consistency for repeated post-disturbance trajectory evaluation.
\item We propose a normalized multi-phase resilience metric that separates disturbance, degraded, and restoration phases. The metric is further embedded in a security-constrained reconnection screening procedure to compare restoration timing and load-ramping strategies.
\end{itemize}


\section{Problem Formulation and Framework Overview}
This paper considers the utility-side assessment of a large data center load.
The utility evaluates grid-facing active and reactive power changes at the
interconnection point without requiring internal models of IT workloads,
server power management, UPS operation, or backup generation.
\subsection{Utility-Side Data Center Disturbance Setting}
\subsubsection{Grid Dynamic Model}
The grid-side post-disturbance dynamics are modeled by nonlinear DAEs formulated as \cite{yan2019fast}
\begin{equation}
    \begin{aligned}
        \dot{\mathbf{y}} &= f(\mathbf{y},\mathbf{z}), & \qquad \mathbf{y}(t_0) &= \mathbf{y}_0, \\
        0 &= g(\mathbf{y},\mathbf{z}), & \qquad \mathbf{z}(t_0) &= \mathbf{z}_0,
    \end{aligned}
    \label{eq:Basic DAE}
\end{equation}
where $\mathbf{y}=\mathbf{y}(t) \in \mathbb{R}^{n_y}$ denotes dynamic states such as frequency, and $\mathbf{z}=\mathbf{z}(t) \in \mathbb{R}^{n_z}$ denotes algebraic states such as voltage. Here, $f(\cdot)$ and $g(\cdot)$ represent the differential and algebraic equations, respectively. In this paper, the DAEs are assumed to be index-1\cite{roche1989implicit}, which is standard in power-system dynamic modeling and sufficient for constructing the proposed PINN-based trajectory predictor.

\begin{figure*}
    \centering
    \includegraphics[trim=10 15 760 900, clip ,width=1\linewidth]{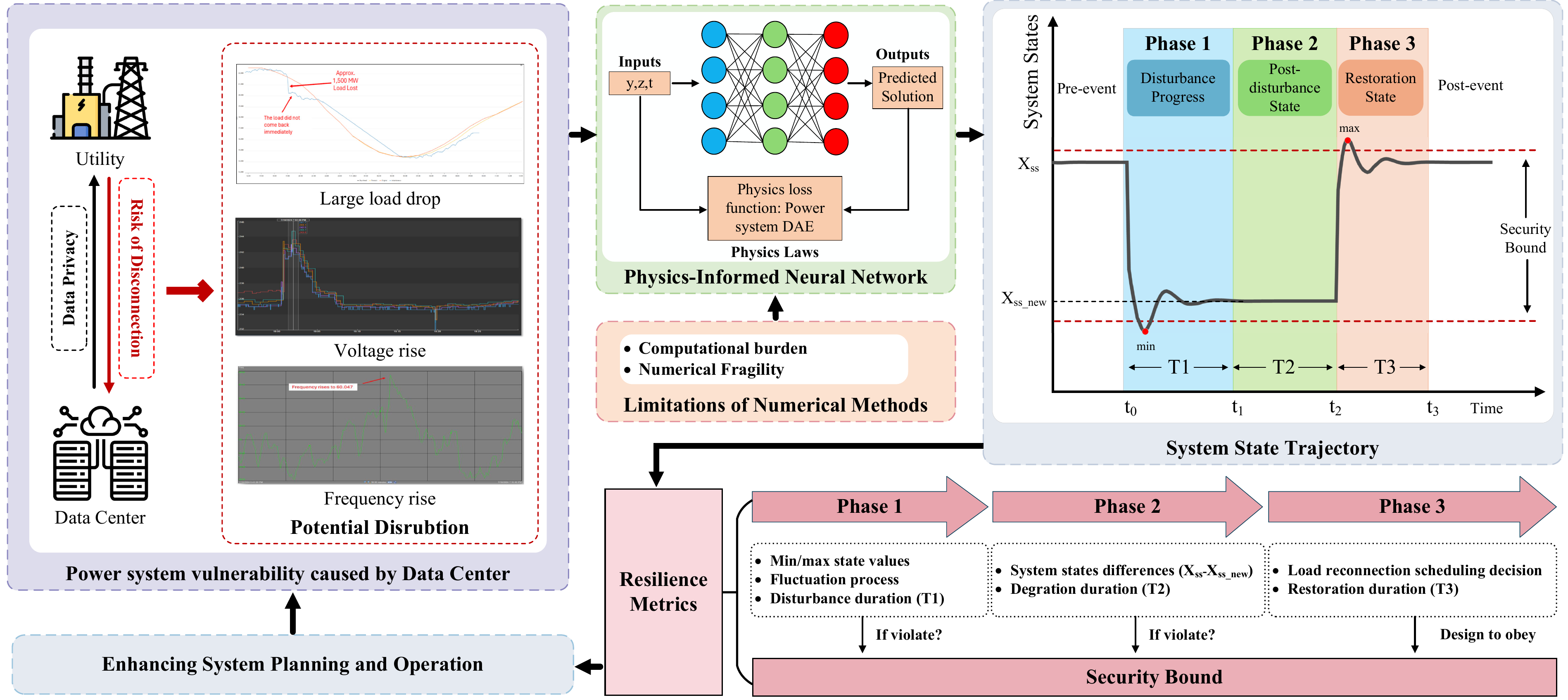}
    \caption{Overview of the proposed resilience assessment framework. Data center disconnection can induce large post-disturbance deviations in system states. A PINN is used to predict the resulting trajectories, which are then evaluated by phase-based resilience metrics for the disturbance, degraded, and restoration phases. The figures in potential disruption can be found in \cite{NERC2025LargeLoadLoss}.}
    \label{fig:Overall diagram}
\end{figure*}

\subsubsection{Trajectory-Based Resilience Screening Problem}

Let \(X_p\), \(X_{\rm topo}\), and \(X_r\) denote system parameters,
network topology, and the restoration decision, respectively. The
post-disturbance trajectory and resilience metric are written as
\begin{equation}
    \mathcal{T}=F(X_p,X_{\rm topo},X_r), \qquad
    R=G(\mathcal{T}).
    \label{eq:trajectory_metric_mapping}
\end{equation}
Here, \(F(\cdot)\) is the trajectory generator and \(G(\cdot)\) is the
trajectory-to-metric map. This formulation highlights the two requirements
of the proposed framework: efficient repeated trajectory generation and
trajectory-level resilience quantification. The response is divided into
disturbance, degraded, and restoration phases, with event times
\(t_0,t_1,t_2,t_3\), as shown in Fig.~\ref{fig:Overall diagram}.

\begin{figure}[h]
    \centering
    \includegraphics[trim=10 15 300 300, clip ,width=0.75\linewidth]{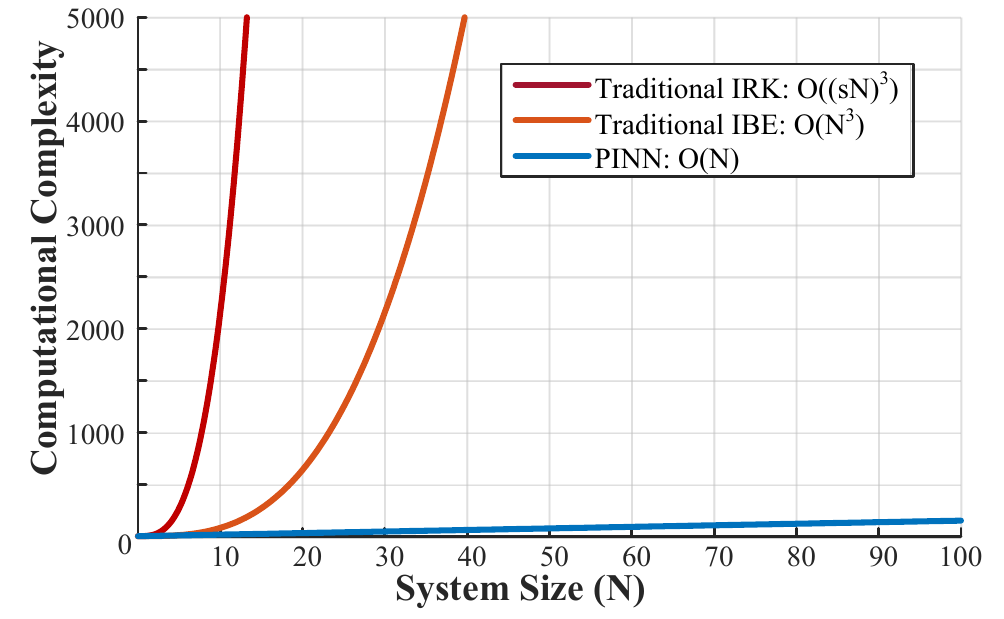}
\caption{Conceptual comparison of repeated numerical DAE simulation and
trained PINN rollout. Implicit solvers require step-by-step nonlinear solves,
whereas a trained PINN uses fixed forward passes after offline training. }
    \label{fig:numerical_vs_PINN}
\end{figure}

\subsection{Need for Efficient Trajectory Generation}

The trajectory-based formulation requires repeated post-disturbance
trajectory generation over many candidate restoration decisions. Numerical
DAE solvers provide reliable references for individual scenarios, but their
step-by-step nonlinear solves become costly when many reconnection times,
load steps, and holding durations must be screened. A trained PINN shifts
this burden to an offline training stage; after training, each rollout is
obtained through neural-network forward passes. Therefore, the PINN is not
introduced to replace numerical solvers as accuracy references, but to
provide a fast physics-consistent trajectory generator for repeated
resilience screening, as conceptually illustrated in
Fig.~\ref{fig:numerical_vs_PINN}.

\subsection{Challenges in PINN-Based Power-System DAE Prediction}

Although PINN-based surrogates are attractive for repeated trajectory
generation, applying them to power-system DAEs is not straightforward. A
power-system dynamic trajectory must satisfy both the differential equations
of dynamic states and the algebraic equations imposed by network power-flow
constraints. Therefore, a PINN used in this paper should not only provide
fast trajectory rollout, but also preserve algebraic consistency and avoid
spurious convergence during training.

\subsubsection{Algebraic Constraint Consistency}
A key challenge in PINN-based dynamic prediction is the
treatment of algebraic constraints. In some existing formulations, the
physics-informed loss is mainly imposed on the differential equations, while
algebraic variables are either ignored, approximated separately, or recovered
through an additional numerical power-flow or simulation procedure
\cite{shen2025physics, stiasny2021transient}. As a result, these models do
not directly act as full DAE trajectory predictors, and their predicted
trajectories may drift from the algebraic manifold when operating conditions
or contingencies change. Although some studies jointly enforce differential
and algebraic residuals, their demonstrations are often limited to small
systems or task-specific state selections \cite{moya2023dae}. For the
resilience screening problem considered in this paper, the trajectory
predictor must output both dynamic and algebraic states while explicitly
penalizing violations of the DAE residuals.

\subsubsection{Training Robustness}

Another challenge is training robustness. Power-system DAEs are nonlinear
and can lead to nonconvex, ill-conditioned physics-informed loss landscapes.
Consequently, a PINN may converge to a solution that has a small residual
loss but does not reproduce the physically correct trajectory. This issue is
particularly important for discrete-time PINN formulations, where the
residual structure depends on the selected integration scheme and time step.
When the time step is small, some residual terms may become weakly scaled,
which can make the optimization problem sensitive to initialization,
loss-weight selection, and the relative scaling between differential and
algebraic residuals.

These challenges motivate the unsupervised DAE-PINN in
Section~III. The proposed formulation uses a single-stage implicit residual
based on backward Euler discretization and jointly predicts dynamic and
algebraic variables. This design keeps the residual dimension compact while
explicitly enforcing both differential and algebraic consistency during
training.

\section{Unsupervised PINN for Post-Disturbance Trajectory Prediction}

\subsection{Power-System DAE and Implicit Residual Formulation}

In this paper, we follow the system DAEs in \cite{zheng2010bi}. 
Assuming that bus $0$ is the slack bus, and $\Omega_{\mathrm{B}}$, $\Omega_{\mathrm{PV}}$, and $\Omega_{\mathrm{PQ}}$ denote bus set, PV bus set, and PQ bus set respectively.
Let us start by defining active and reactive power residuals $R_{P_i}$ and $R_{Q_i}$:

\begin{equation}
    \label{eq:active power}
    R_{P_i} =  U_i \sum_{j=1}^{n} U_j \left( G_{ij} \cos \delta_{ij} + B_{ij} \sin \delta_{ij} \right) + P_i^{l} - P_i^{g}
\end{equation}
\begin{equation}
    \label{eq:reactive power}
    R_{Q_i} =  U_i \sum_{j=1}^{n} U_j \left(G_{ij} \sin \delta_{ij} - B_{ij} \cos \delta_{ij} \right) + Q_i^{l} - Q_i^{g}
\end{equation}
where $P_i^g$ ($P_i^l$) and $Q_i^g$ ($Q_i^l$) denote the active and reactive power of generators (loads) at bus $i$, and if it is a PQ bus, the generation power will be 0.
$U_i$ and $\delta_i$ represent the voltage and phase angle of bus $i$.
$j$ is the neighbors of bus $i$, $\delta_{ij}:= \delta_i - \delta_j$ denotes the voltage phase angle
difference between buses $i$ and $j$, 
and 
$G_{ij}$ and
$B_{ij}$ are the real and imaginary parts of the admittance matrix element
$Y_{ij}$. 

Let \(d\) denote the data center interconnection bus. From the utility side,
the data center is modeled as a grid-facing constant-power load component
\((P_{\rm dc},Q_{\rm dc})\) at bus \(d\). Its disconnection and reconnection
are represented by an availability factor \(\ell(t)\in[0,1]\):
\begin{equation}
    P_d^l(t)=P_{d,0}^l+\ell(t)P_{\rm dc},\qquad
    Q_d^l(t)=Q_{d,0}^l+\ell(t)Q_{\rm dc},
    \label{eq:dc_load_model}
\end{equation}
where \(P_{d,0}^l\) and \(Q_{d,0}^l\) denote the non-data-center load at bus \(d\). 
The availability factor is the normalized form of the reconnection load \(P(t)\) in Section~IV-B, i.e., \(P(t)=\ell(t)P_{\max}\), where
\(P_{\max}\) denotes the full data-center loading level. Equivalently, when
\(P(t)\) is expressed in active-power units, \(P_{\max}=P_{\rm dc}\), and the
same factor \(\ell(t)\) scales \(Q_{\rm dc}\) under a fixed data-center power
factor. Thus, before disconnection \(\ell(t)=1\); after disconnection
\(\ell(t)=0\); and during reconnection \(\ell(t)\) follows the normalized
step-ramping profile.

\textit{Remark 1}: In this work, the data center is represented as a grid-facing large load change rather than a detailed facility-side dynamic model. This choice is consistent with the utility-side assessment setting considered in this paper: the dominant externally visible impact of a disconnection event is an abrupt reduction in net demand at the interconnection bus. The model is therefore intended to capture the grid-side voltage and frequency response to large load loss and reconnection, while detailed internal dynamics such as UPS controls, workload migration, and backup generation are outside the scope of this study.

Then the power system DAEs can be formulated as:
\begin{align}
\label{eq: slack bus dynamics}
\dot{\omega}_0
&= \frac{1}{M_0}(-D_0\omega_0 + \sum_i R_{P_i}),
&& i \in \Omega_B \setminus \{0\} \\
\dot{\omega}_i
&= \frac{1}{M_i}\!\left(-D_i\omega_i - R_{P_i}\right),
&& i \in \Omega_{PV} 
\label{eq: PV_omega}\\
\dot{\delta}_i
&= \omega_i - \omega_0,
&& i \in \Omega_{PV} 
\label{eq: PV_delta}\\
\dot{\delta}_i
&= -\omega_0 - \frac{R_{P_i}}{D_i},
&& i \in \Omega_{PQ} 
\label{eq: PQ_delta} \\
0 &= -\frac{1}{U_i}R_{Q_i}, && i \in \Omega_{PQ} 
\label{eq: PQ_algebraic}
\end{align}
where $\mathbf{y} = (\omega_0, \omega_i, \delta_i, \delta_j)^{\mathsf T}$ for $i \in\Omega_{PV}, j\in\Omega_{PQ}$ denotes the
vector of dynamic states, and $\mathbf{z}=(U_i)_{i\in\Omega_{PQ}}$ is the vector of algebraic states. $M$ is the inertia
constants, and $D$ represent the damping coefficients.

Assume that the dynamic states and algebraic variables at time $t_n$ are $(\mathbf{y}_n, \mathbf{z}_n)$ separately.
To advance it to $(\mathbf{y}_{n+1}, \mathbf{z}_{n+1})$ at $t_{n+1}$, we apply the implicit Backward Euler \cite{biswas2013discussion} methods to solve DAEs~(\ref{eq: slack bus dynamics}) - (\ref{eq: PQ_algebraic}) and obtain:
\begin{equation}
    \label{eq:backward Euler}
    \begin{aligned}
    &\mathbf{y}_{n+1} - \mathbf{y}_n - h f(\mathbf{y}_{n+1}, \mathbf{z}_{n+1}) = 0,\\
    &g(\mathbf{y}_{n+1}, \mathbf{z}_{n+1}) = 0
    \end{aligned}
\end{equation}
where $t_{n+1} = t_n + h$ and $h>0$ is the time step, $f(\cdot)$ represents the differential equations, and $g(\cdot)$ are the algebraic equations. In a conventional implicit solver, (\ref{eq:backward Euler}) would be solved by Newton iterations at each time step. In the proposed PINN, the same implicit residual is instead used as the physics-informed training objective, as described in Section III-C.


\subsection{PINN Architecture}
\begin{figure}
    \centering
    \includegraphics[trim=10 10 490 680, clip, width=1.0\linewidth]{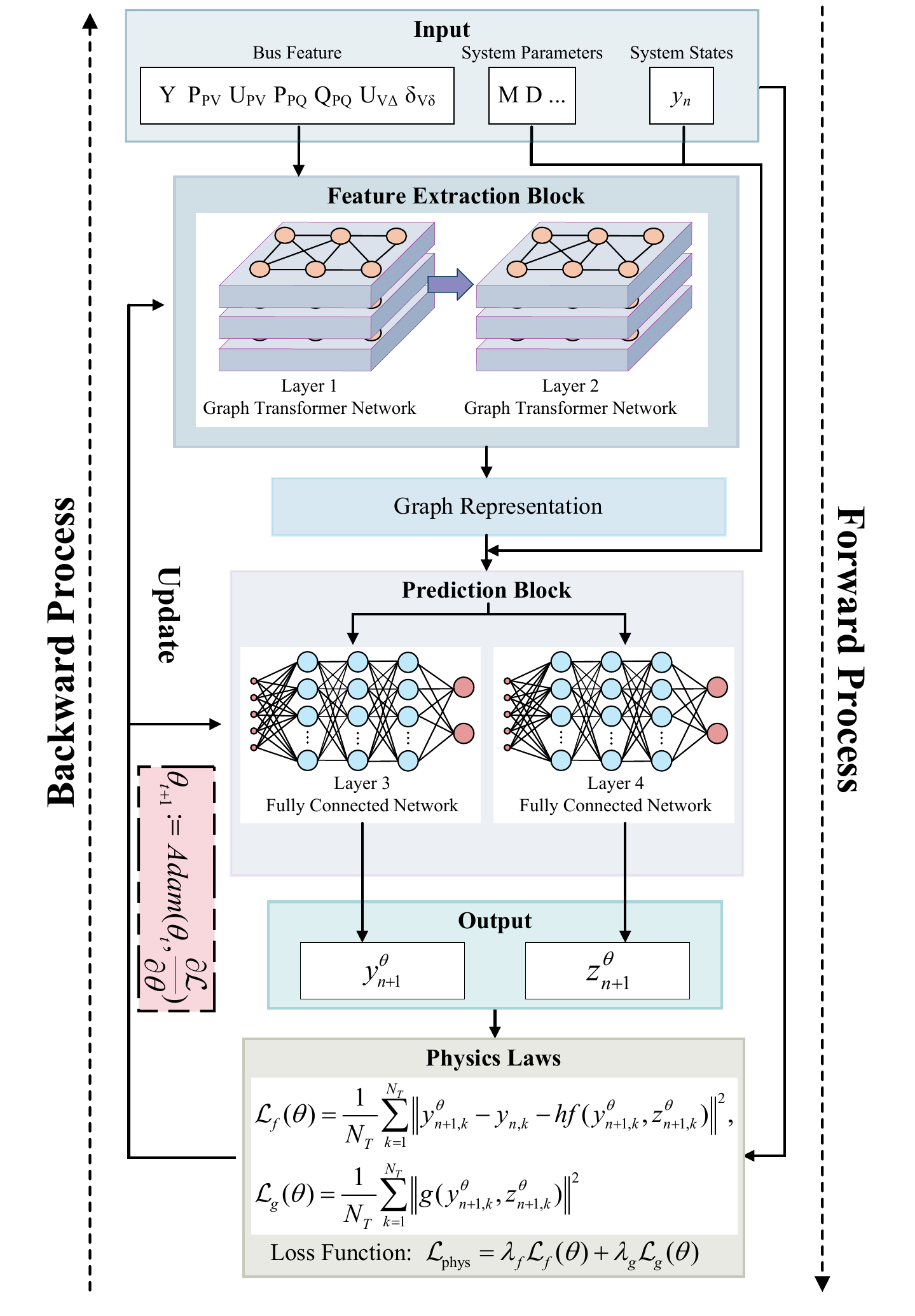}
    \caption{Model architecture of the employed PINN }
    \label{fig:NN architecture}
\end{figure}
The graph-based feature extraction block is used to encode network topology and bus-level electrical features before next-step state prediction. This design allows the model input to reflect both local bus attributes and electrical-neighborhood interactions, which are important for voltage and frequency propagation after large load changes.
Fig.~\ref{fig:NN architecture} illustrates the overall PINN architecture used in this paper.
The inputs of network $\textbf{X} $ can be divided into three parts $(\textbf{X}_1, \textbf{X}_2,\textbf{X}_3)$. The first one  $\textbf{X}_1$ includes admittance matrix $Y$, the active and reactive power of PQ buses, the active
power and voltage magnitudes of PV buses, and the voltage magnitude and phase angle of the slack bus.
$\textbf{X}_2$ represents system parameters $(M, D)$, and $\textbf{X}_3$ is the current system states $\mathbf{y}_n$.
The PINN model has two main parts: feature extraction block and prediction block.
The feature extraction block consists of two Graph Attention Network (GAT) layers~\cite{velivckovic2017graph} and takes $\mathbf{X}_1$ as its input:
The bus-admittance matrix $\mathbf{Y}$ defines the graph adjacent matrix $A$, while the remaining components are used as node attributes. For bus $i$, a GAT layer updates its representation as
\begin{equation}
\mathbf{x}'_i
=
f_\sigma\left(
\sum_{j\in\mathcal{N}(i)\cup\{i\}} A_{ij}
\alpha_{ij}\,\boldsymbol{\Theta}\mathbf{x}_j
\right),
\label{eq:GAT_update}
\end{equation}
where $\mathcal{N}(i)$ denotes the electrical neighborhood of bus $i$, 
$\boldsymbol{\Theta}$ is a learnable weight matrix, and $f_\sigma(\cdot)$ is the non-linear activation function.
$\mathbf{x}_i$ and $\mathbf{x}_j$ denote the input feature vectors of buses $i$ and $j$, while $\mathbf{x}'_i$ is the corresponding
output embedding of bus $i$ after attention-based aggregation and the nonlinear layer.
The attention coefficients $\alpha_{ij}$ are computed as
\begin{equation}
\alpha_{ij}
=
\mathrm{softmax}_{j}\!\left(
\mathrm{LeakyReLU}\!\left(
\mathbf{a}^\top
\big[
\boldsymbol{\Theta}\mathbf{x}_i
\Vert
\boldsymbol{\Theta}\mathbf{x}_j
\big]
\right)
\right),
\label{eq:GAT_attention}
\end{equation}
where $\mathbf{a}$ is a trainable attention vector. By stacking two such layers, the feature extraction
block produces the graph representation matrix $\mathbf{H}$.
This graph representation is then flattened and concatenated with system parameters $\mathbf{X}_2$ and system states $\mathbf{X}_3$ to form the input of the prediction block:
\begin{equation}
\mathbf{X}^{p} = 
\left[
flatten(\mathbf{H}) \Vert \mathbf{X}_2 \Vert \mathbf{X}_3
\right].
\label{eq:pred_input}
\end{equation}
The prediction block adopts a multi-task learning structure: two parallel fully connected networks are employed to separately predict system dynamic states $\mathbf{y}_{n+1}$ and algebraic variables $\mathbf{z}_{n+1}$ at the next time step:
\begin{align}
\mathbf{y}_{n+1} &= \mathrm{FC}_y(\mathbf{X}^{p}) =
f_\sigma\!\left(
\mathbf{W}^{y} \mathbf{X}^{p} + \mathbf{b}^{y}
\right), 
\label{eq:pred_y}\\
\mathbf{z}_{n+1} &=\mathrm{FC}_z(\mathbf{X}^{p}) =
f_\sigma\!\left(
\mathbf{W}^{z} \mathbf{X}^{p} + \mathbf{b}^{z}
\right),
\label{eq:pred_z}
\end{align}
where $\mathbf{W}$ and $\mathbf{b}$ are weight and
bias coefficients.
Then the output $\mathbf{y}_{n+1}$, $\mathbf{z}_{n+1}$, and network input $\textbf{X}$ together are sent to the physics loss function. 




\subsection{Unsupervised PINN Loss Function}
In this paper, we adopt an unsupervised PINN formulation based on the implicit backward Euler discretization of the power-system DAEs in (\ref{eq:backward Euler}):
\begin{equation}
\label{eq:whole loss function}
    \mathcal{L}_{\text{phys}} = \lambda_f \mathcal{L}_f(\theta) + \lambda_g \mathcal{L}_g(\theta)
\end{equation}
where
\begin{align}
    \mathcal{L}_f(\theta) &=  \frac{1}{N_T}\sum_{k=1}^{N_T} \left\| \mathbf{y}_{n+1, k}^{\theta} - \mathbf{y}_{n, k} - h f(\mathbf{y}_{n+1, k}^{\theta}, \mathbf{z}_{n+1, k}^{\theta}) \right\|^2,
    \label{eq:dynamic_loss}\\
    \mathcal{L}_g(\theta) &=  \frac{1}{N_T}\sum_{k=1}^{N_T} \left\| g(\mathbf{y}_{n+1, k}^{\theta}, \mathbf{z}_{n+1, k}^{\theta}) \right\|^2
\end{align}
where $\mathbf{y}_{n, k}$ is the input to the PINN, $\mathbf{y}_{n+1, k}^{\theta}$ and $\mathbf{z}_{n+1, k}^{\theta}$ are the predicted outputs characterized by $\theta$, and $N_T$ is the number of training samples. Instead of matching labeled trajectories, the model is trained directly through the discrete physics. This loss jointly enforces the differential and algebraic equations and therefore avoids the mismatch between learned dynamics and algebraic consistency.

In implementation, the residual components are scaled by characteristic magnitudes of the corresponding states and algebraic variables to reduce imbalance between frequency, angle, and voltage residuals. The weights $\lambda_f$ and $\lambda_g$ are then used only to balance the differential and algebraic residual groups, rather than to compensate for unit differences among individual state variables.

\begin{figure}[t]
    \centering
    \begin{subfigure}[b]{0.49\linewidth}
        \centering
        \includegraphics[trim=10 170 25 180, clip,width=\linewidth]{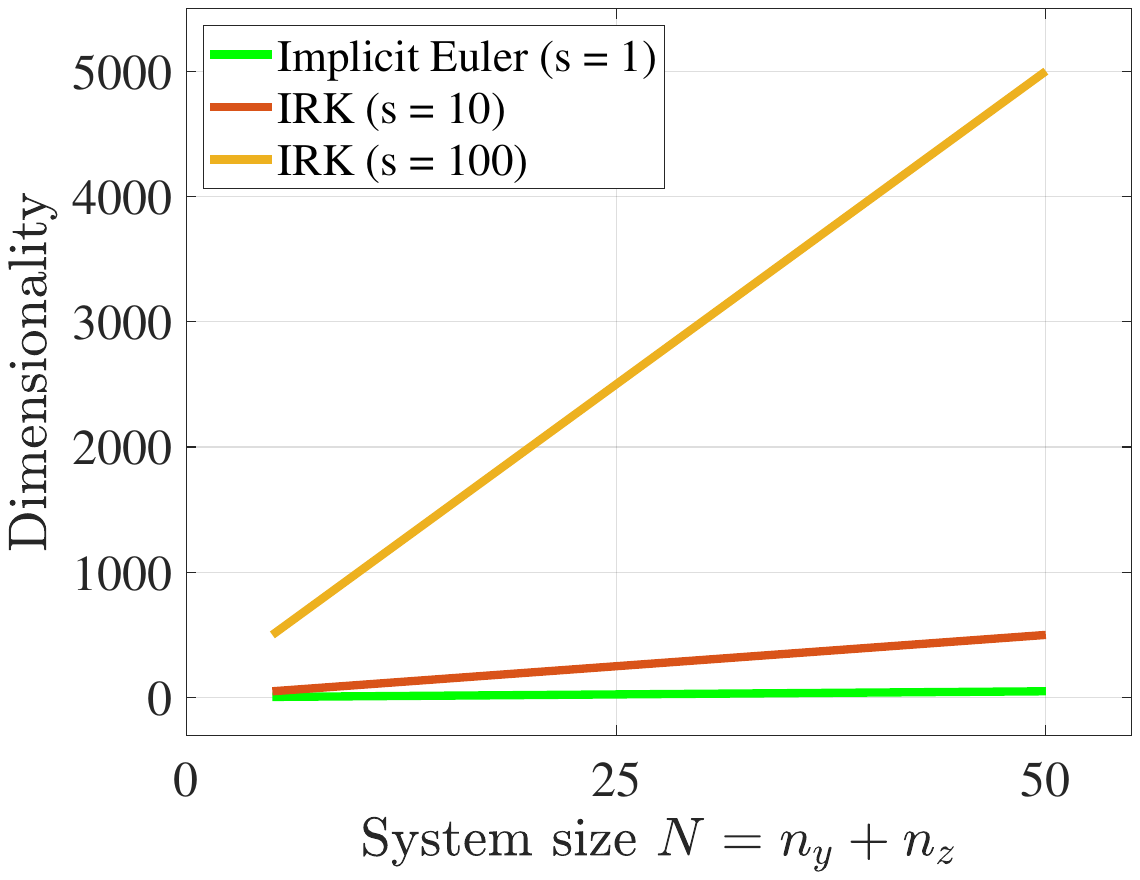}
        \caption{Loss dimension}
        \label{fig:dimension explosion}
    \end{subfigure}
    \hfill
    \begin{subfigure}[b]{0.49\linewidth}
        \centering
        \includegraphics[trim=20 170 25 180, clip,width=\linewidth]{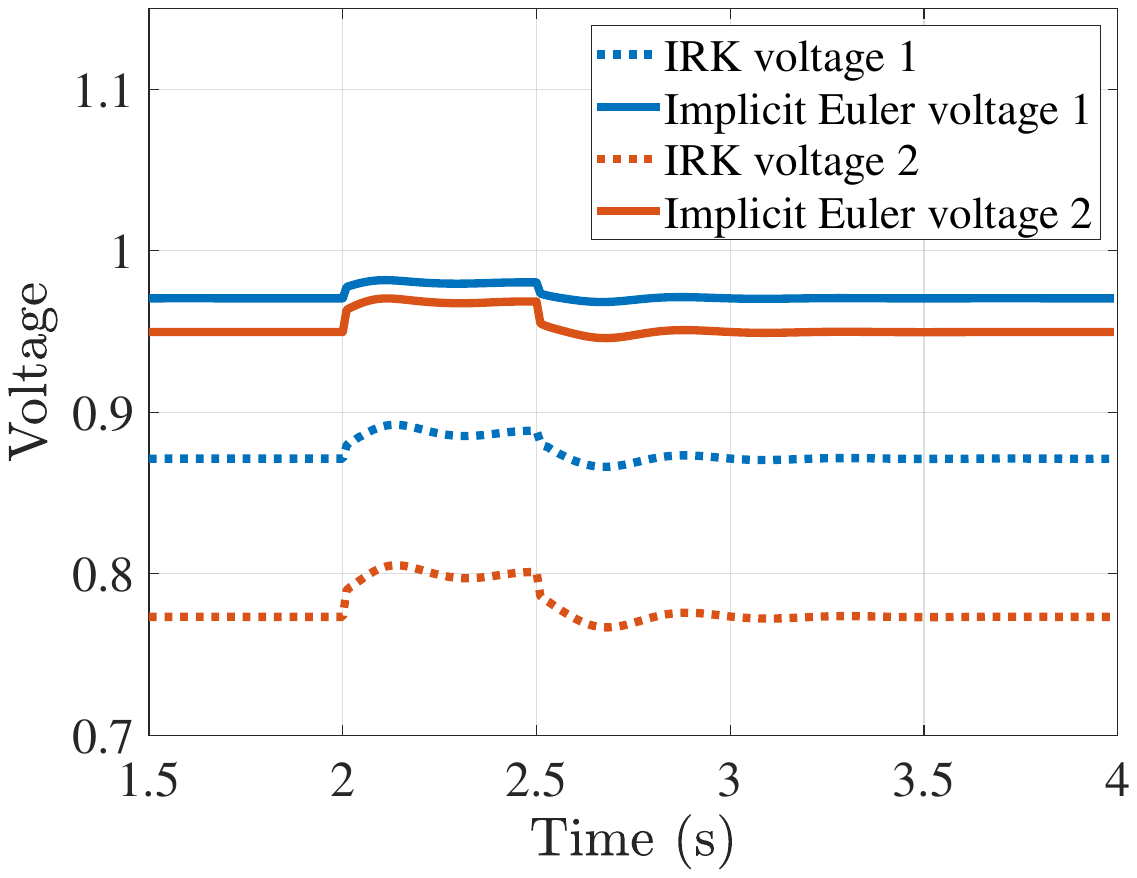}
        \caption{Voltage trajectories }
        \label{fig:voltage difference}
    \end{subfigure}
    \caption{Comparison of optimization burden and voltage trajectories for Implicit Euler and IRK. (a) The loss dimension increases with the number of internal stages $s$, making IRK harder to optimize.(b) For small $h$, IRK converges to an incorrect low-voltage solution, while Implicit Euler tracks the correct high-voltage solution.}
    \label{fig:IRK_IBE}
\end{figure}

The PINN in \cite{moya2023dae} also follows an unsupervised, equation-driven strategy, but its loss is built on an implicit Runge--Kutta formulation. For an $s$-stage IRK method, the network must output stage variables $\bigl(\mathbf{y}_k^{(i)}, \mathbf{z}_k^{(i)}\bigr)_{i=1}^s$, so the number of unknowns and residual terms per time step scales as $(n_y+n_z)s$. As system size grows, and as large $s$ is used to maintain long-horizon accuracy, memory usage, Jacobian size, and automatic differentiation cost increase substantially. 
In addition, the loss contains $n_y s$ differential constraints per time step, and when the time step $h$ is small, all the corresponding physics gradients vanish with $h$, creating a high-dimensional, competing loss landscape that is difficult to optimize reliably. 
In our experiments, this high-dimensional residual structure made training more sensitive to initialization and step-size selection. For small step sizes, the IRK-based PINN was observed to converge to an incorrect low-voltage solution branch in the tested system.
By contrast, the proposed unsupervised PINN uses one implicit stage per time step. Therefore, the number of predicted next-step variables and residual terms per sample scales with $n_y+n_z$ and the number of differential constraints is reduced to $n_y$ per step. This reduces the dimension of the physics-informed loss and improves training robustness in the tested post-disturbance trajectory prediction tasks.
Fig. \ref{fig:IRK_IBE} illustrates this effect: as in Fig. \ref{fig:dimension explosion}, IRK methods with large stage numbers rapidly increase the loss dimension, while Implicit Euler remains compact. Fig. \ref{fig:voltage difference} shows the consequence at small step sizes, where IRK collapses to the low-voltage branch but Implicit Euler remains on the correct high-voltage manifold.

\section{Multi-Phase Resilience Metrics And Security-Constrained Screening}
This section defines multi-phase resilience metrics for data center
disconnection and reconnection events. The metrics quantify accumulated
state deviations during disturbance, degraded, and restoration phases. Larger metric values indicate larger accumulated deviation from desired operating conditions and therefore lower resilience.

\subsection{Phase-Wise Resilience Metric Formulation}

The disturbance and reconnection events follow the utility-side load model
in \eqref{eq:dc_load_model}.
Let the system state at time $t$ be $
(\mathbf{U}_t,\boldsymbol{\delta}_t,\boldsymbol{\omega}_t)$,
where $\mathbf{U}_t\in\mathbb{R}^{N_u}$, $\boldsymbol{\delta}_t\in\mathbb{R}^{N_\delta}$, and
$\boldsymbol{\omega}_t\in\mathbb{R}^{N_\omega}$ denote the \textit{\textbf{per-unit}} bus voltage magnitudes, phase angles, and frequency deviations, respectively.
For each phase $p$ with time interval $[t_{\mathrm{s}},t_{\mathrm{e}}]$,
define a state-wise resilience functional
\begin{equation}
r_p^{k} = \mathcal{M}_p\!\left(\mathbf{x}^{k}_{t}\right),
\qquad k\in\{U,\delta,\omega\},t \in [t_{\mathrm{s}},t_{\mathrm{e}}]
\label{eq:statewise_metric_def}
\end{equation}
where $\mathbf{x}^{U}_{t}=\mathbf{U}_t$, $\mathbf{x}^{\delta}_{t}=\boldsymbol{\delta}_t$,
$\mathbf{x}^{\omega}_{t}=\boldsymbol{\omega}_t$, and $r_p^{k}\in \mathbb{R}^{N_k} $ is a vector, computed element-wise over different buses and phases.
$\mathcal{M}_p(\cdot)$ is the chosen phase-specific metric; consider three phases disturbance process $[t_0,t_1]$, degraded state $[t_1,t_2]$,
and restorative phase $[t_2,t_3]$. For each state component $k$,
define the phase-wise metrics:

\textbf{(i) Disturbance process:} system fluctuation and disturbance duration
\begin{equation}
\mathcal{M}_1(\mathbf{x}^{k}_t) = \int_{t_0}^{t_1}
\left|\mathbf{x}^{k}_t-\mathbf{x}^{k}_{\mathrm{ss}}\right|\,\mathrm{d}t.
\label{eq:r1}
\end{equation}
where $x_{ss}^{k}$ denotes the pre-disturbance nominal steady-state value, $t_0$ represents when the disturbance occurs, and $t_1$ is the timestamp at which the system reaches a new quasi-steady degraded state. 
This metric captures both the whole fluctuation process and the disturbance duration ($t_1 - t_0$), quantifying the system’s absorptive capacity and resistance to the onset of data center-induced disturbances.

\textbf{(ii) Degraded state:} steady-state deviation and degraded duration
\begin{equation}
\mathcal{M}_2(\mathbf{x}_t^k)=
|\mathbf{x}_{ss,d}^k-\mathbf{x}_{ss}^k|(t_2-t_1)_+,
(t_2-t_1)_+=\max(t_2-t_1,0).
\label{eq:r2}
\end{equation}
where $\mathbf{x}_{\mathrm{ss,d}}^{k}$ denotes the degraded system equilibrium.
In this formulation, $|x_{ss\_d}^k - x_{ss}^k|$ measures the steady-state value deviation, and 
the term $(t_2 - t_1)$ defines the degraded state duration, which is the window during which the system remains at its lowest performance level before active recovery begins.
$\max(\cdot, 0)$ acts as a saturation function to evaluate the impact of operator-decided restoration time $t_2$. It will be discussed in the later subsection.

\textbf{(iii) Restorative phase:} time to recovery and restoration response
\begin{equation}
\mathcal{M}_3(\mathbf{x}^{k}_t)  = \int_{t_2}^{t_3}
\left|\mathbf{x}^{k}_t-\mathbf{x}^{k}_{\mathrm{ss\_p}}\right|\,\mathrm{d}t.
\label{eq:r3}
\end{equation}
Here, $t_3$ denotes the recovery completion time, and $\mathbf{x}^{k}_{\mathrm{ss\_p}}$ denotes the target post-restoration steady state, which may differ from the pre-disturbance steady state. After computing the bus-wise vectors, each metric is aggregated into a scalar score,
\begin{equation}
\bar{r}_p^{k} = \mu\!\left(\mathbf{r}_p^{k}\right),
\qquad k\in\{U,\delta,\omega\},
\label{eq:bus_mean}
\end{equation}
where $\mu(\cdot)$ is an aggregation operator, chosen as the mean in this paper. The phase-level resilience metric is then defined as the weighted sum over states:
\begin{equation}
R_p = \beta^U\bar r_p^{U} + \beta^\delta\bar r_p^{\delta} + \beta^\omega\bar r_p^{\omega}.
\label{eq:phase_resilience}
\end{equation}
where $\beta^k$ represent the state $k$ weight.

The overall system-level resilience metric is defined as the weighted sum of phase-level resilience metrics:
\begin{equation}
R_{\mathrm{sys}} = \sum_{p=1}^{3} \alpha_pR_p=\alpha_1 R_1 + \alpha_2 R_2 + \alpha_3 R_3,
\label{eq:system_resilience}
\end{equation}
where $\alpha_p \ge 0$ and $\alpha_1+\alpha_2+\alpha_3=1$.

\subsection{Security-Constrained Reconnection Screening}
After a data center disturbance, a candidate reconnection strategy is
defined as \(X_r=(t_2,\Delta P,\Delta T_{\rm hold})\), where \(t_2\) is the
restoration initiation time, \(\Delta P\) is the load step, and
\(\Delta T_{\rm hold}\) is the holding time between reconnection actions.
Each strategy is screened by its restoration metric \(R_3(X_r)\), recovery
duration \(T_{\rm rec}=t_3-t_2\), and voltage/frequency violation durations
\(d_v(X_r)\). The feasible strategy set is
\begin{equation}
    \mathcal{X}_{\rm feas}=
    \{X_r\in\mathcal{X}_r: d_v(X_r)\leq \bar d_v,\ \forall v\in\mathcal{V}\},
\end{equation}
where \(\bar d_v\) is the allowable duration for violation type \(v\).
Feasible strategies are then compared using the trade-off between
\(R_3(X_r)\) and \(T_{\rm rec}\).

\subsubsection{Restoration Timing Selection}

\begin{figure}
    \centering
    \begin{subfigure}[b]{0.49\linewidth}
        \centering
        \includegraphics[trim=20 480 775 150, clip,width=\linewidth]{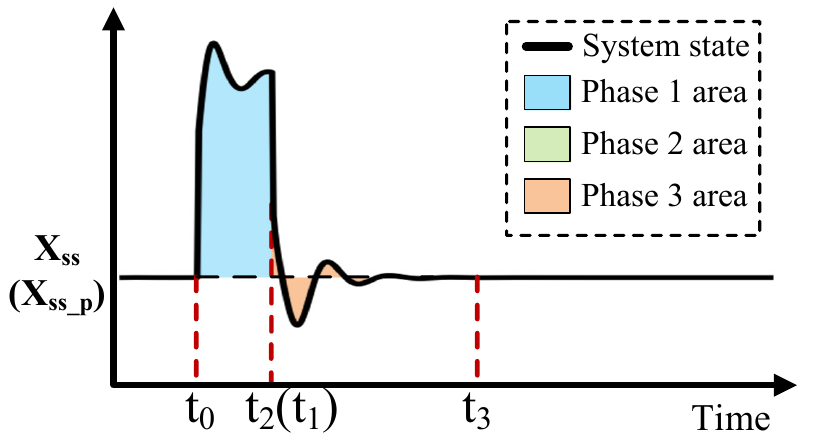}
        \caption{$t_1 > t_2$}
        \label{fig:t1 large than t2}
    \end{subfigure}
    \hfill
    \begin{subfigure}[b]{0.49\linewidth}
        \centering
        \includegraphics[trim=20 230 775 410, clip,width=\linewidth]{Visio-t2.pdf}
        \caption{$t_1 < t_2$}
        \label{fig:t1 smaller than t2}
    \end{subfigure}
    \caption{Impact of $t_2$ values on resilience metrics. (a)  $t_2 < t_1$: restoration begins before the degraded state is reached, shortening Phase 1, eliminating Phase 2, and advancing the start of Phase 3. (b) $t_2 > t_1$: restoration begins after the degraded state is reached, producing a nonzero Phase-2 metric.}
    \label{fig:t2}
\end{figure}

The restoration initiation time \(t_2\) changes the phase durations. If
\(t_2\leq t_1\), the degraded phase is eliminated and \(M_2=0\) through the
positive-part term in \(M_2\). If \(t_2>t_1\), the system remains in the
degraded state for \(t_2-t_1\), increasing Phase-2 resilience loss. Fig.~\ref{fig:t2}
illustrates these two cases.

\subsubsection{Load Ramping Strategy Selection}
Beyond timing, the choice of load ramping pattern also significantly impacts the system response $x_t$ during Phase 3.
The load reconnection ramping pattern considered in this study is based on the reconnection loading process specified in Dominion Energy Virginia’s Facility Interconnection Requirements\cite{DominionEnergyVirginia2025FIR}.
It can be formulated as:
\begin{equation}
\label{eq: load ramp}
    P(t) = \min \left( \left( \left\lfloor \frac{t - t_{2}}{\Delta T_{\text{hold}}} \right\rfloor + 1 \right) \Delta P, \ P_{\text{max}} \right),\; t\geq t_2.
\end{equation}
where $P(t)$ is the data center load at time $t$.
$\Delta T_{\text{hold}}$ is the hold duration, which represents the time interval between successive reconnection actions.
The magnitude of each step load ramping is controlled by $\Delta P$: a larger $\Delta P$ value means larger step increasing magnitude.
The floor function $\lfloor \cdot \rfloor$ forces the output to stay at a constant integer for the entire duration of $\Delta T_{\text{hold}}$.
The $+1$ offset ensures that the restoration begins with an immediate load injection at $t_{\text{2}}$. The $\min(\cdot)$ saturation operator locks the load parameter at the maximum capacity $P_{\text{max}}$ once the staircase reaches the prescribed target.

\subsubsection{Restoration Security Constraint Check}
To ensure that a selected restoration time and load ramping strategy is feasible, security checks are incorporated into the Phase 3 assessment. Table \ref{tab:violation_limits} gives the maximum allowable violation durations for different severity levels under IEEE 1547-2018 \cite{basso2014ieee}. For any restoration strategy, the resulting transient trajectory $x_t$ must satisfy these limits. If any violation duration exceeds its prescribed bound, the corresponding
strategy is classified as infeasible and excluded from
\(\mathcal{X}_{\rm feas}\).

\begin{table}[t]
\centering
\caption{Security Violation Limits and Allowed Duration}
\label{tab:violation_limits}
\renewcommand{\arraystretch}{1.5}
\begin{tabular}{lcc}
\hline
\textbf{Violation Severity} & \textbf{Bound} & \textbf{Allowed Duration} \\
\hline
Severe Over-voltage & $U > 1.20$ pu & $0.16$ s \\
Moderate Over-voltage  & $U > 1.10$ pu & $2.00$ s \\
Moderate Under-voltage  & $U < 0.70$ pu & $2.00$ s \\
Severe Under-voltage & $U < 0.45$ pu & $0.16$ s \\
Frequency &\rule{0pt}{15pt}\makecell[l]{$f > 62.0$ Hz \\$f < 56.5$ Hz}   & $0.16$ s \\
\hline
\end{tabular}
\end{table}

\section{Numerical Study} 
\subsection{Experiment Setups}
\begin{figure}[h]
    \centering
    \includegraphics[trim=20 10 250 355, clip,width=0.8\linewidth]{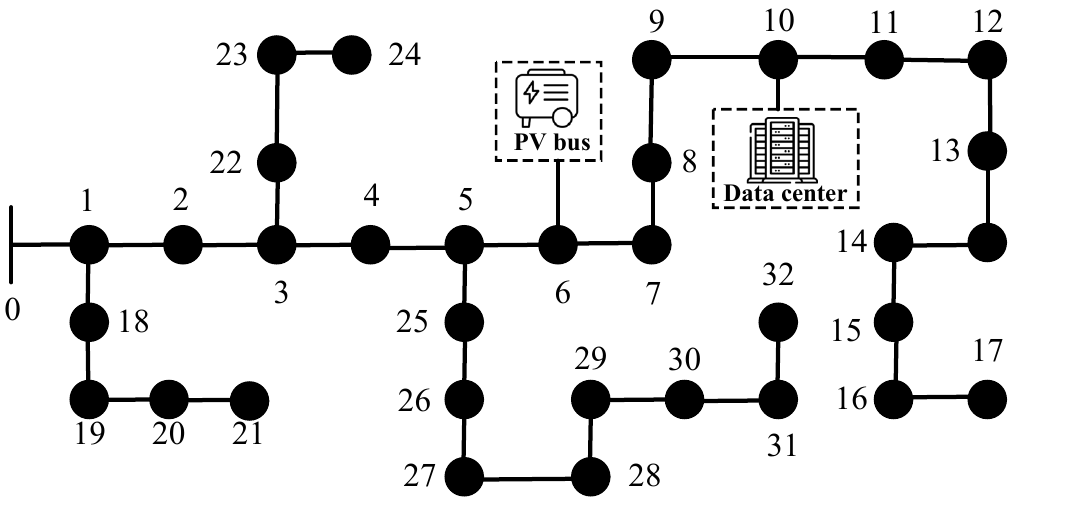}
    \caption{Modified IEEE 33-bus test feeder used in the case study. Bus 10 is selected as the PV bus and bus 6 is selected as the data center load bus. Extensions to multiple large-load buses and larger networks follow the same DAE formulation but are left for future validation.
    }
    \label{fig:IEEE_33}
\end{figure}

The numerical study evaluates three aspects of the proposed framework: (i) the trajectory prediction accuracy and algebraic consistency of the DAE-PINN, (ii) the ability of the normalized phase-wise metrics to distinguish disturbance severity and location, and (iii) the computational benefit of PINN-enabled repeated screening of reconnection strategies.
As is shown in Fig.~\ref{fig:IEEE_33}, bus 10 is selected as a PV bus and bus 6 as the data center load bus so that the system has 34 dynamic states and 31 algebraic variables as defined in Eq.(\ref{eq: slack bus dynamics})-Eq. (\ref{eq: PQ_algebraic}).
For PINN training and testing, the Adam optimizer with default hyperparameters is used with an initial learning rate $lr=10^{-3}$, which is reduced linearly during training. The training and test datasets are generated from randomly sampled initial conditions
$\omega_i(0) \sim \mathcal{U}(-\pi,\,\pi),$ and $
\delta_i(0) \sim  \mathcal{U}(-0.5,0.5).$
The time step is set to $h=0.01$. In the loss function (Eq.~\ref{eq:whole loss function}), the weights are chosen as $\lambda_f=100$ and $\lambda_g=1$ to compensate for the larger magnitude of the algebraic loss $\mathcal{L}_g(\theta)$ compared with the dynamic loss $\mathcal{L}_f(\theta)$, thereby avoiding compromising gradient optimization.
The GNN layers and dynamic-state output layer use the $\mathrm{SiLU}(\cdot)$ activation function, while the algebraic output layer adopts $\mathrm{softplus}(\cdot)$. 
The models are implemented in PyTorch and trained on an NVIDIA GeForce RTX 4070 GPU.

\subsection{PINN Model Results}
We first verify the effectiveness of the PINN in performing a long-time simulation of DAEs.
We recurrently update (in a Markov chain style) the input to the PINN, $\mathbf{y}_{n+1}$, using the predicted dynamic states $\mathbf{y}^{\theta}_{n+1}$ from the previous evaluation step.
Then in the time interval $[0, h\cdot N]$, the system state trajectory is obtained by concatenating the outputs from all forward passes, i.e., $\{Y_k^{\theta}\}_{k=1}^{N}$ and $\{Z_k^{\theta}\}_{k=1}^{N}$. 
Here, the state trajectories are evaluated for $N=800$ time steps with a time step size of $h=0.01$. 

\begin{figure}
    \centering
    \includegraphics[trim=30 40 170 70, clip,width=0.9\linewidth]{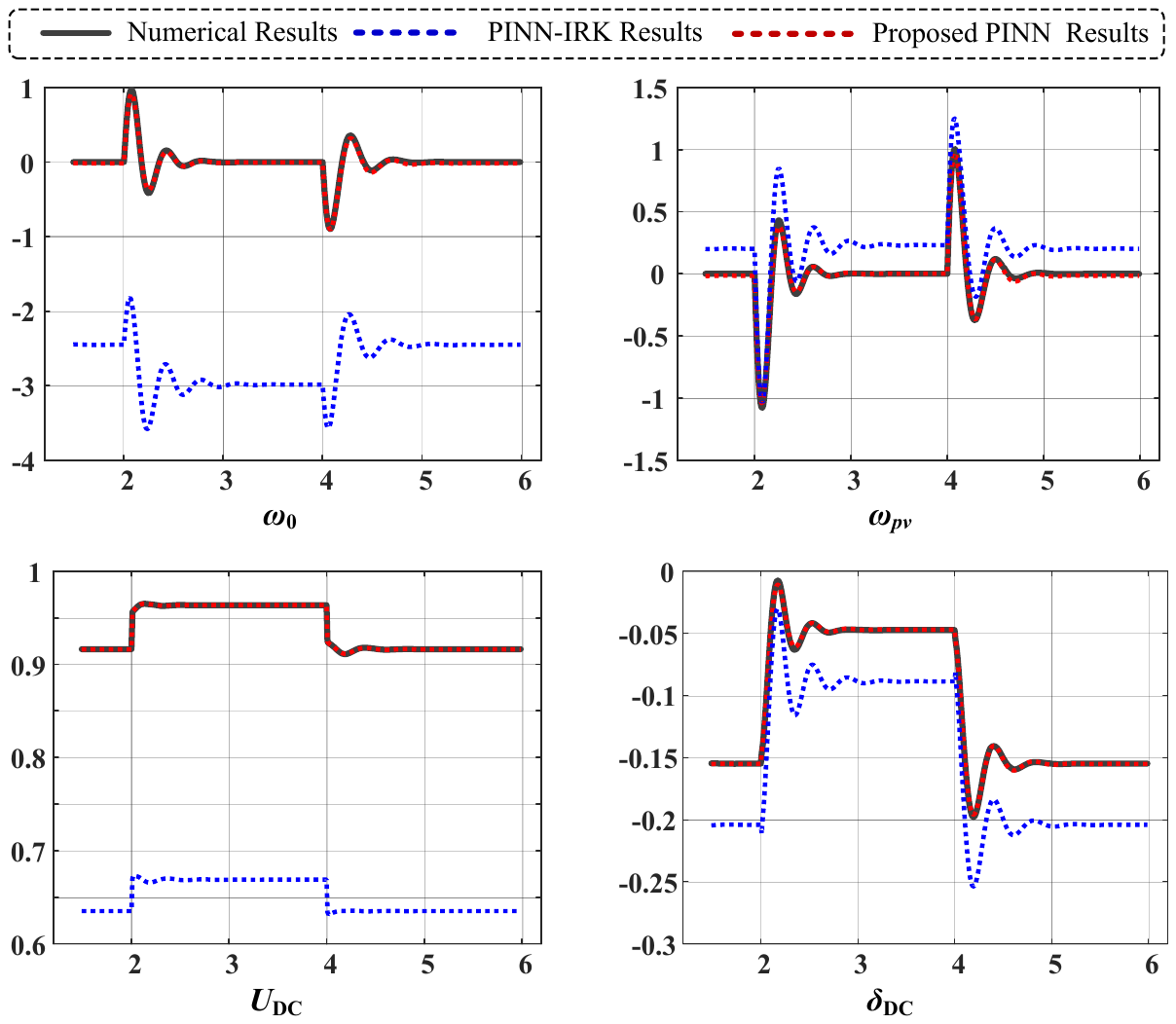}
    \caption{Comparison of state trajectories obtained from the numerical DAE solver, the PINN-IRK approach, and the proposed PINN method for representative dynamic and algebraic state trajectories. The displayed states include angular frequencies for slack bus $\omega_0$ and PV bus $\omega_{PV}$, voltage magnitudes for data center bus $U_{DC}$ and its phase angle $\delta_{DC}$. The proposed PINN closely matches the numerical solver, whereas PINN-IRK shows large deviations.}
    \label{fig:PINN vs numerical}
\end{figure}

Fig.~\ref{fig:PINN vs numerical} compares the trajectories predicted by the proposed PINN and the PINN-IRK approach against the reference numerical DAE solver. The selected plots show representative system variables, including the frequencies of the slack bus and PV bus, and the voltage magnitude and phase angle at the data center bus. As shown in the figure, the proposed PINN closely matches the numerical solution throughout the entire simulation horizon, capturing both transient oscillations and steady-state values. In contrast, the PINN-IRK method exhibits significant deviations, especially in voltage trajectories, which is consistent with the convergence issues discussed earlier.


To further quantify the predictive accuracy, Table \ref{tab:dataset_stats} reports the mean square error (MSE) and standard deviation between the PINN outputs and the numerical benchmark for all system states. The errors remain very small across frequency, phase-angle, and voltage variables, which confirms that the proposed PINN maintains high accuracy over long simulation horizons.

\begin{table}
\centering
\caption{
MSE and standard deviation of the proposed PINN model predictions relative to the numerical solver for different system states ($\omega_i$, $\delta_i$, $U_i$) and algebraic constraints error $\|g(\mathbf{y}^\theta,\mathbf{z}^\theta)\|_2$ over the long-time simulation.
}
\begin{tabular}{lcccc}
\toprule
 & $\omega_i$ & $\delta_i$ & $U_i$ & $\|g(\mathbf{y}^\theta,\mathbf{z}^\theta)\|_2$ \\
\midrule
MSE     & 4.59e-04  & 7.57e-07  & 7.78e-08 & 4.40018e-05 \\
St.\ dev. & 1.00e-03 &  2.49e-06 & 5.41e-07 & 2.22176e-06\\
\bottomrule
\end{tabular}
\label{tab:dataset_stats}
\end{table}

\subsection{Metric Response to Disturbance Severity and Location}
A comprehensive analysis was performed to compare the resilience metrics across two scenarios: different data center load level and data center positions within the system.
For simplicity, we set $\alpha_p$ in Eq.(\ref{eq:system_resilience}) all to be 1/3 to obtain system resilience metric $R_{sys}$, and we choose direct reconnection strategy in restoration phase to focus on metrics performance.
$t_0$ and $t_2$ are set to be 2s and 4s, respectively.
The experimental results are as follows.

\subsubsection{Different Data Center Load Levels}
We first compare resilience metrics across phase and system level under different data center load levels. The data center is set at bus 6.
Table~\ref{tab:resilience_metrics_load_level} reports the resilience metrics under different data center load levels of 20\%, 30\%, and 40\% of the total load. As the load increases, metrics $R_1$, $R_2$, and $R_3$ all increase, leading to a larger overall resilience metric $R_{\mathrm{sys}}$. Since these metrics quantify cumulative deviation from steady-state behavior, larger values indicate greater resilience loss. This trend is consistent with the expected physical response: a larger load loss produces larger voltage, angle, and frequency deviations and therefore larger accumulated normalized deviation. The result confirms that the proposed metric is directionally consistent with disturbance severity.

\begin{table}[h]
\centering
\caption{Resilience metrics under different load levels}
\label{tab:resilience_metrics_load_level}
\begin{tabular}{c|ccc|c}
\toprule
Load percentage & $R_1$ & $R_2$ & $R_3$ & $R_{sys}$ \\
\midrule
40\% & 0.2703 & 0.1484 & 0.1981 & 0.20560 \\
30\% & 0.1952 & 0.1137 & 0.1389 & 0.14956 \\
20\% & 0.1264 & 0.0766 & 0.0881 & 0.09704 \\
\bottomrule
\end{tabular}
\end{table}

\begin{figure}
    \centering
        \begin{subfigure}[b]{0.93\linewidth}
        \centering
        \includegraphics[trim=20 270 25 660, clip,width=\linewidth]{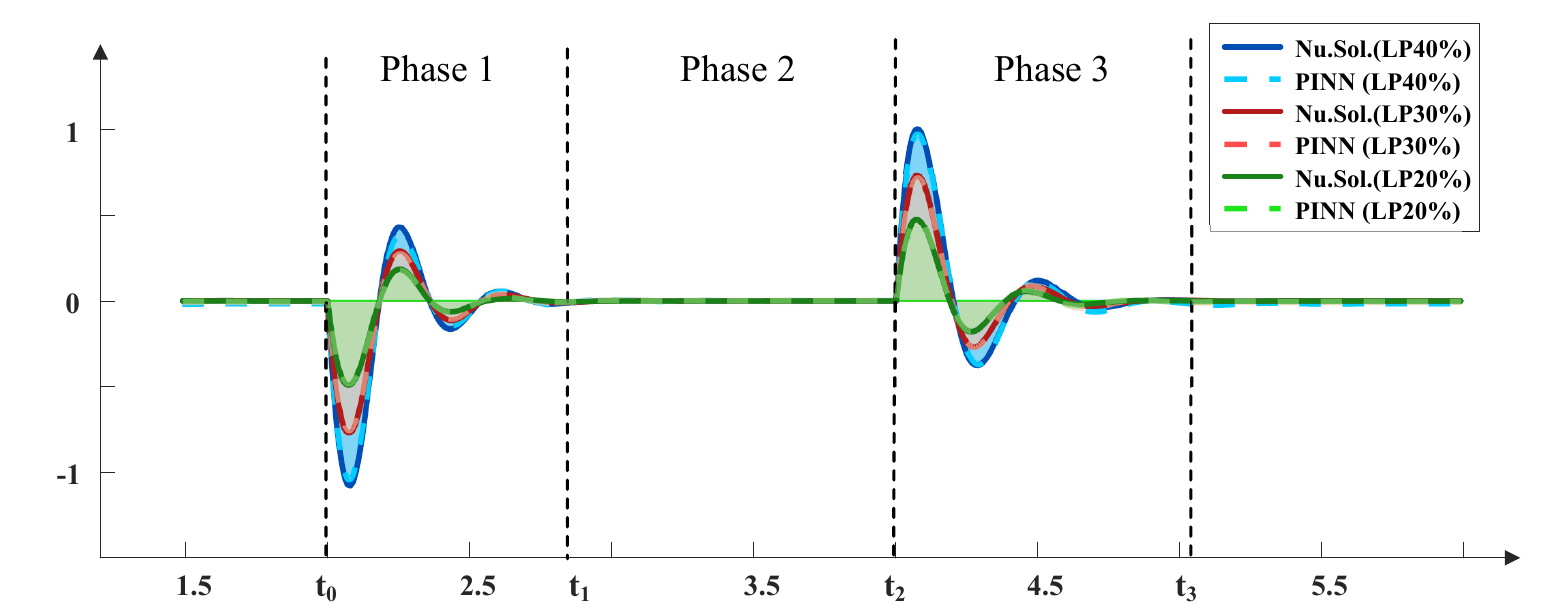}
        \caption{Frequency trajectories at slack bus }
        \label{fig:voltage}
    \end{subfigure}
        \begin{subfigure}[b]{0.93\linewidth}
        \centering
        \includegraphics[trim=20 870 25 60, clip,width=\linewidth]{Visio-load.pdf}
        \caption{Voltage trajectories at data center load bus}
        \label{fig:dimension}
    \end{subfigure}
    \hfill
    \begin{subfigure}[b]{0.93\linewidth}
        \centering
        \includegraphics[trim=20 570 25 360, clip,width=\linewidth]{Visio-load.pdf}
        \caption{Phase angle trajectories at data center load bus}
        \label{fig:phase_angle}
    \end{subfigure}
    \caption{System trajectories from the Numerical Solver (solid lines) and PINN predictions (dashed lines) under 20\%, 30\%, and 40\% load levels. Subplots show (a) slack-bus frequency, (b) data-center-bus voltage, and (c) data-center-bus angle over the three phases. 
    Colored areas indicate the metric calculation regions for each load level (blue for 40\%, red for 30\%, and green for 20\%).
    We can see that higher load increases all resilience metrics, indicating weaker system resilience.
    }
    \label{fig:load percentage}
\end{figure}

Fig.~\ref{fig:load percentage} shows representative system trajectories under different data center load percentages. The 40\% load case exhibits the largest oscillations and deviation areas during both the disturbance and restoration phases, which is consistent with the larger metric values in Table~\ref{tab:resilience_metrics_load_level}. The close agreement between the numerical solutions and PINN predictions again demonstrates the accuracy of the proposed trajectory predictor across different disturbance severities.

\subsubsection{Different Data Center Locations}

The second scenario considers different locations of the data center load within the network. Specifically, buses 2, 6, and 15 are selected to represent locations near the front, middle, and leaf portions of the feeder. Table~\ref{tab:dc_location} shows that the resilience metrics increase as the data center location moves farther from the slack bus. When the disturbance occurs close to the slack bus, the resulting state deviations are smaller and the resilience loss is lower. In contrast, disturbances at electrically weaker or more remote locations lead to larger deviations across all phases. These results show that the proposed metrics can effectively capture the influence of network location on system resilience.

\begin{figure}[h]
    \centering
        \begin{subfigure}[b]{0.93\linewidth}
        \centering
        \includegraphics[trim=20 290 40 650, clip,width=\linewidth]{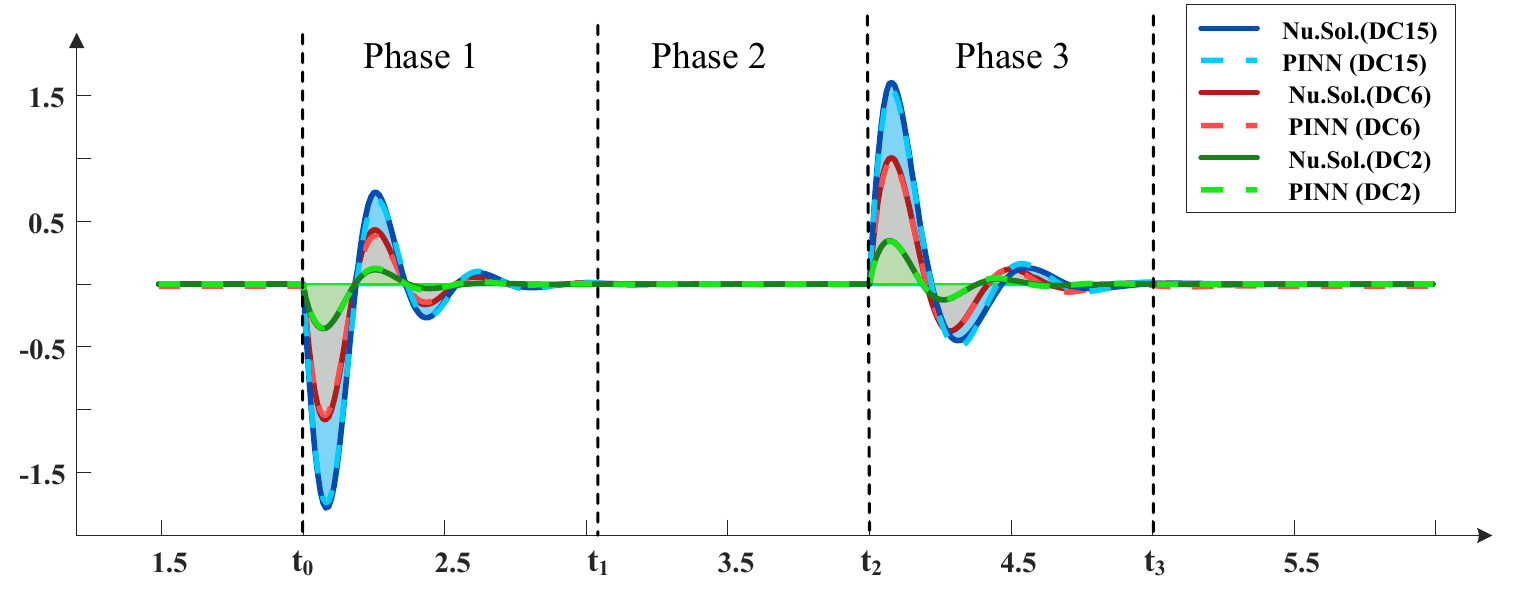}
        \caption{Frequency trajectories at slack bus }
        \label{fig:DC position freq}
    \end{subfigure}
        \begin{subfigure}[b]{0.93\linewidth}
        \centering
        \includegraphics[trim=20 880 40 60, clip,width=\linewidth]{Visio-dc_position.pdf}
        \caption{Voltage trajectories at data center load bus}
        \label{fig:DC position dimension}
    \end{subfigure}
    \begin{subfigure}[b]{0.93\linewidth}
        \centering
        \includegraphics[trim=20 580 40 360, clip,width=\linewidth]{Visio-dc_position.pdf}
        \caption{Phase angle trajectories at data center load bus}
        \label{fig:DC position voltage}
    \end{subfigure}
    \caption{System trajectories from the Numerical Solver (solid lines) and PINN predictions (dashed lines) for data center locations at bus 15, bus 6, and bus 2. Subplots show (a) slack-bus frequency, (b) data-center-bus voltage, and (c) data-center-bus angle over the three phases. 
    Colored areas indicate the resilience metric calculation regions for each load level (blue for bus 15, red for bus 6, and green for bus 2).  As the data center moves farther from the slack bus, all resilience metrics increase, indicating lower system resilience.}
    \label{fig:DC position}
\end{figure}

Fig.~\ref{fig:DC position} illustrates representative trajectories for different data center locations. When the data center is placed closer to the slack bus, the disturbance produces smaller deviations in system states. As the location moves farther along the feeder, the oscillations and deviation areas become larger. The close tracking between the numerical solver and the PINN prediction again confirms the accuracy of the proposed framework under different disturbance locations.

\begin{table}[h]
\centering
\caption{Resilience metrics for different data center locations}
\label{tab:dc_location}
\begin{tabular}{c|ccc|c}
\toprule
Data center position & $R_1$ & $R_2$ & $R_3$ & $R_{sys}$ \\
\midrule
DC at bus 2  & 0.0957 & 0.05678 & 0.0612 & 0.07127 \\
DC at bus 6  & 0.2703  & 0.1484  & 0.1981  & 0.20560  \\
DC at bus 15 & 0.4619 & 0.2281  & 0.3166  & 0.33556 \\
\bottomrule
\end{tabular}
\end{table}


The results in Table~\ref{tab:PV distance} also indicate that, for the
tested feeder and control settings, the resilience metric is affected not
only by the distance to the slack bus but also by the location relative to
the PV bus. When $d_{\rm slack}$ is fixed, the cases closer to the PV bus
show larger resilience loss. A possible explanation is that load
disconnection near a voltage-regulated bus more directly perturbs the local
active and reactive power balance and causes stronger interaction with the
PV-bus voltage constraint. This interpretation is system-dependent and
should be further examined in larger networks and with multiple generation
locations.


\begin{table}
\centering
\caption{Resilience metrics for different $d_{slack}$ and $d_{PV}$}
\label{tab:PV distance}
\begin{tabular}{l|cc|ccc|c}
\toprule
DC location & $d_{slack}$ & $d_{PV}$ & $R_1$ & $R_2$ & $R_3$ & $R_{\text{sys}}$ \\
\midrule
DC at bus 6  & 6 & 4  & 0.2703 & 0.1484 & 0.1981 & 0.2056 \\
DC at bus 25 & 6 & 6  & 0.2558 & 0.1479 & 0.1835 & 0.1957 \\
\hline
DC at bus 2  & 2 & 8  & 0.0957 & 0.0568 & 0.0612 & 0.0712 \\
DC at bus 18 & 2 & 10 & 0.0521 & 0.0356 & 0.0309 & 0.0395 \\
\bottomrule
\end{tabular}
\end{table}

\subsection{Restoration Strategy Evaluation}
We next illustrate how the proposed metrics can be used to compare data center load restoration strategies. In this experiment, the data center load percentage is fixed at 40\%.

\subsubsection{Load-Ramping Pattern Selection}

\begin{figure}[h]
    \centering
    \begin{subfigure}[b]{0.49\linewidth}
        \centering
        \includegraphics[trim=20 600 665 520, clip,width=\linewidth]{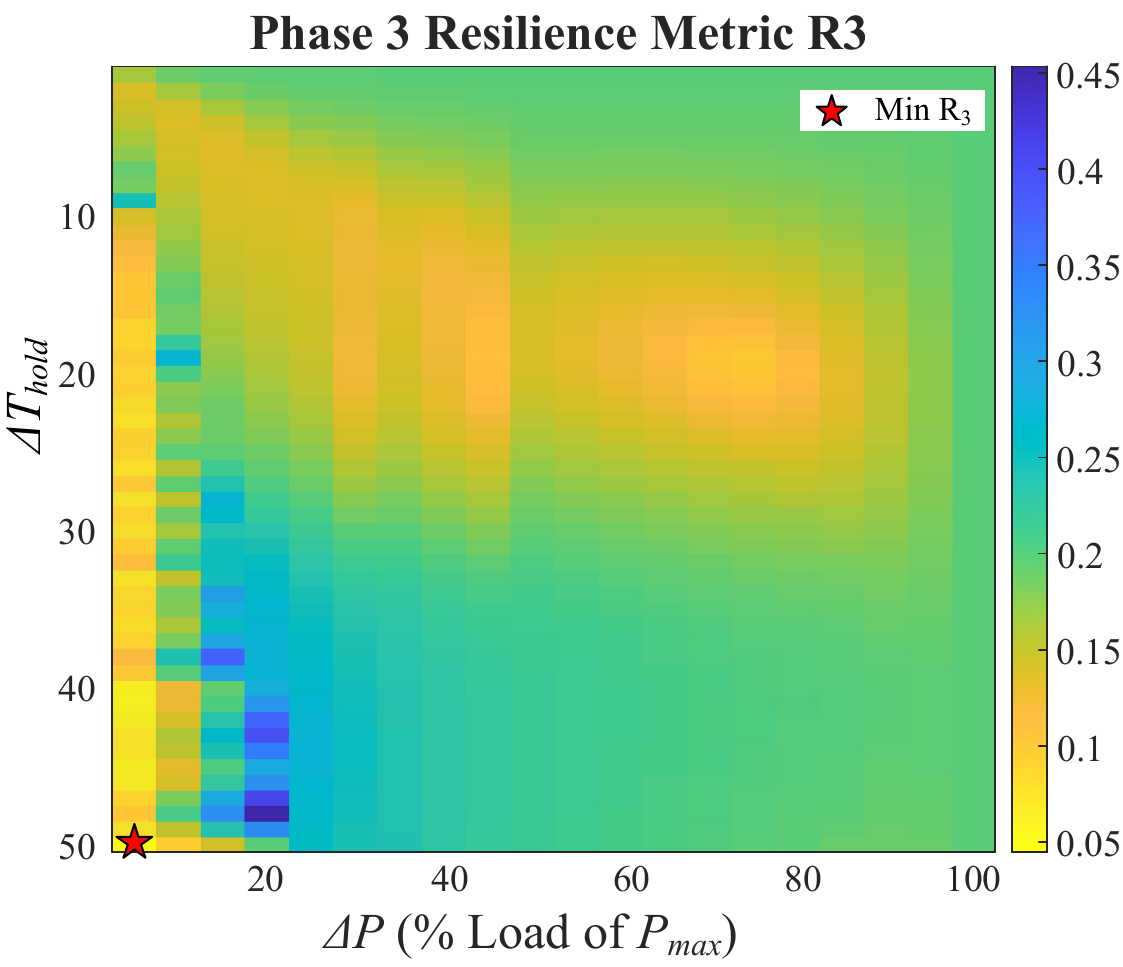}
        \caption{$R_3$ under different $\Delta P$ and $\Delta T_{\text{hold}}$ }
        \label{fig:R3 comparison}
    \end{subfigure}
    \begin{subfigure}[b]{0.49\linewidth}
        \centering
        \includegraphics[trim=570 600 110 520, clip,width=\linewidth]{Visio-heat_maps.pdf}
        \caption{Recovery duration under different $\Delta P$ and $\Delta T_{\text{hold}}$}
        \label{fig:duration comparison}
    \end{subfigure}
    \begin{subfigure}[b]{0.85\linewidth}
        \centering
        \includegraphics[trim=60 20 150 990, clip,width=\linewidth]{Visio-heat_maps.pdf}
        \caption{Comparison of $R_3$ and recovery duration for different load ramping patterns}
        \label{fig:R3vsduration}
    \end{subfigure}
    \caption{Effect of load ramping parameters on Phase-3 restoration. (a) Resilience deviation metric $R_3$ for different $\Delta P$ and $\Delta T_{\text{hold}}$, where smaller increments and longer holding times reduce deviation. (b) Phase-3 recovery duration for different $\Delta P$ and $\Delta T_{\text{hold}}$, where larger increments and shorter holding times accelerate recovery. (c) Trade-off between recovery duration and system deviation, with highlighted points marking the fastest-recovery and minimum-deviation strategies.}
    \label{fig:different load restoration}
\end{figure}

\begin{figure}
    \centering
    \includegraphics[trim=30 240 25 35, clip,width=0.85\linewidth]{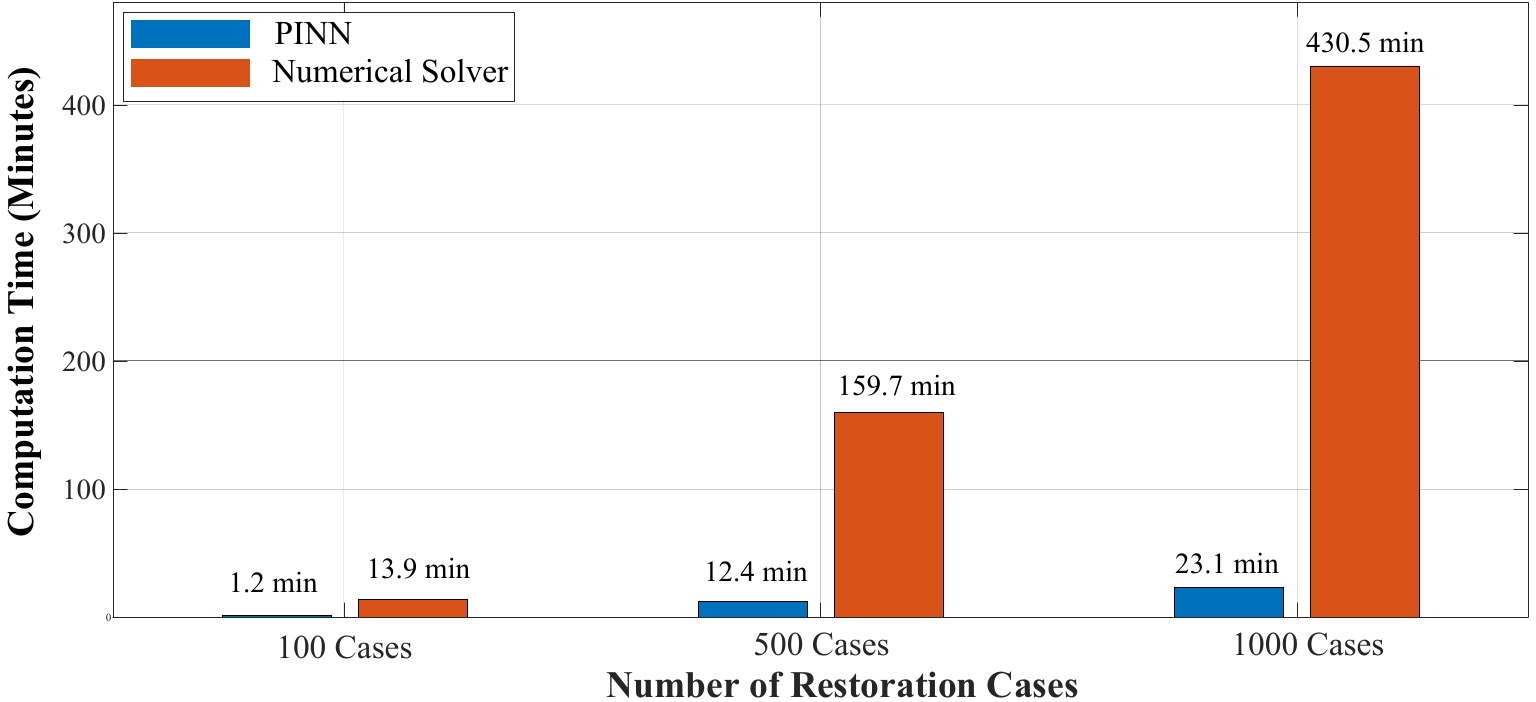}
    \caption{Computation time comparison between the proposed PINN framework and the numerical DAE solver. }
    \label{fig:computation cost}
\end{figure}

To analyze the effect of the two main restoration parameters $\Delta P$ and $\Delta T_{\text{hold}}$, we perform a two-dimensional grid search. Specifically, $\Delta P$ ranges from $5\%$ to $100\%$ of $P_{\max}$ in increments of $5\%$, and $\Delta T_{\text{hold}}$ varies from 1 to 50 time steps. This generates 1000 restoration scenarios. For each scenario, we evaluate the resilience metric $R_3$ and the corresponding recovery duration.
Fig.~\ref{fig:different load restoration} shows a clear trade-off between resilience and restoration speed. Smaller load increments and longer holding intervals generally reduce the Phase-3 resilience loss, but they also increase the recovery duration. In contrast, more aggressive ramping strategies shorten restoration time at the cost of larger transient deviations. The scatter plot in Fig.~\ref{fig:R3vsduration} further confirms that no single strategy can simultaneously minimize both $R_3$ and recovery duration. Therefore, the preferred restoration strategy depends on the operator's objective, that is, whether faster recovery or smaller transient deviation is prioritized. The main computational value of the proposed PINN is not replacing one numerical simulation, but enabling repeated evaluation of many candidate strategies. Therefore, we compare the total computation time required to evaluate 100, 500, and 1000 restoration cases.
Fig.~\ref{fig:computation cost} compares the computation time of the proposed PINN and the numerical DAE solver for computing these restoration cases. For 1000 restoration cases, the proposed PINN requires 23.1 minutes, whereas the numerical DAE solver requires 430.5 minutes. This reduction demonstrates why a trained physics-informed surrogate is useful for repeated restoration screening.

\subsubsection{Restoration-Time Selection}

Another restoration decision variable is the restoration initiation time $t_2$. In principle, earlier restoration reduces the degraded-phase resilience loss. However, in practical grid operation, $t_2$ cannot be arbitrarily small because the system may still be in a fragile transient state immediately after the disturbance. Therefore, the practical choice of $t_2$ should balance prompt recovery and secure reconnection.

\section{Conclusion}

This paper proposed a physics-informed resilience assessment framework for power systems with large data center loads. By developing an unsupervised PINN based on an implicit backward Euler integration scheme, the framework can accurately predict post-disturbance DAE trajectories while explicitly enforcing differential and algebraic consistency. Based on the predicted trajectories, multi-phase resilience metrics were introduced to quantify system performance during the disturbance, degraded, and restoration phases. Case studies on the IEEE 33-bus system showed that the proposed PINN accurately tracks the numerical DAE solver and that the proposed metrics can effectively distinguish the resilience impact of different disturbance severities, disturbance locations, and restoration strategies. The results also showed a clear trade-off between restoration speed and transient resilience performance. Future work will extend the framework to larger systems with multiple data center loads and more detailed dynamic models.






\bibliographystyle{IEEEtran}
\bibliography{bibtex/bib/IEEEexample}
\end{document}